# Warping effects in the band and angular-momentum structures of the topological insulator $Bi_2Te_3$

Wonsig Jung,[1] Yeongkwan Kim,[1] Beomyoung Kim,[1] Yoonyoung Koh,[1] Chul Kim,[1] Masaharu Matsunami,[2] Shinichi Kimura,[2] Masashi Arita,[3] Kenya Shimada,[3] Junghoon Han,[4] Juyoung Kim,[5] Beongki Cho,[5] and Changyoung Kim[1,*]

[1]*Institute of Physics and Applied Physics, Yonsei University, Seoul, Korea*
[2]*UVSOR Facility, Institute for Molecular Science and Graduate University for Advanced Studies, Okazaki 444-8585, Japan*
[3]*Hiroshima Synchrotron Radiation Center, Hiroshima University, Higashi-Hiroshima, Hiroshima 739-0046, Japan*
[4]*Department of Physics, Sungkyunkwan University, Suwon, Korea*
[5]*School of Physics and Department of Materials Science and Engineering, GIST, Gwangju 500-712, Korea*

We performed angle-resolved photoemission (ARPES) experiments on $Bi_2Te_3$ with circularly polarized light. ARPES data show very strong circular dichroism, indicating the existence of orbital angular momentum (OAM). Moreover, the alignment of OAM is found to have a strong binding energy dependence. Such energy dependence comes from a relatively strong band warping effect in $Bi_2Te_3$ compared to $Bi_2Se_3$. OAM close to the Dirac point has an ideal chiral structure ($\sin\theta$) without an out-of-plane component. The warping effect comes in as the binding energy decreases, and circular dichroism along a constant energy contour can no longer be explained by a simple $\sin\theta$ function but requires a $\sin 3\theta$ term. When the warping effect becomes even stronger near the Fermi energy, circular dichroism gains an additional $\sin 6\theta$ term. Such behavior is found to be compatible with the theoretically predicted OAM structure.

Topological insulators (TIs) are a new class of insulators. They are distinguished from ordinary insulators by the $Z_2$ topological number. The most important aspect of such a topological state is the existence of topological metallic (TM) states at the boundary, that is, on the surface. These metallic states have Dirac dispersions and are protected against time-reversal symmetry-conserving perturbations. In addition, the states on the opposite sides of the Dirac cone are time-reversal pairs, resulting in suppression of backscattering. The time-reversal pairs are normally realized in the form of opposite electron spin states due to the Rashba effect from spin-orbit coupling, forming chiral spin states along a constant energy contour.[1–6]

In an ideal case, the Dirac dispersion of TM states is linear and has circular constant energy contours in the momentum space.[7] In such a case, backscattering of the surface electrons by nonmagnetic impurities should be completely suppressed.[8,9] However, angle-resolved photoemission (ARPES) data from $Bi_2Se_3$ and $Bi_2Te_3$ show hexagonal warping of the surface state Fermi surfaces.[10,11] Subsequent theoretical studies reveal that a strong warping effect can modify the spin chirality and brings in an out-of-plane spin polarization component.[12,13] Considering the fact that the spin chirality on the Fermi surface can affect the transport properties, it is desirable to have systematic experimental investigations of the warping effect on the spin states. Unfortunately, detection of the spin state by spin-resolved photoemission is not an easy task due to its low efficiency.

Recent ARPES data that utilized circularly polarized light (CPL) may provide a way to circumvent the problem. The data from $Bi_2Se_3$ show a large circular dichroism (CD), which stems from the existence of orbital angular momentum (OAM).[14] It was found that the OAM is also locked to the electron momentum and its direction is antiparallel to the spin angular momentum (SAM) direction. Therefore OAM also forms a chiral structure and is an essential ingredient previously missing in the description of the surface Dirac states. Such observation provides an opportunity to find the spin direction (by measuring the OAM direction) without spin detectors, which have very low efficiency. One may thus be able to investigate the warping effect on the spin structure in $Bi_2Te_3$ by using dichroic ARPES.

We report on our attempt to investigate the warping effect on the spin structures of the surface Dirac states of $Bi_2Te_3$. Our main goal was to obtain the systematic change in the OAM (hence the SAM, too) structure in the momentum space at a constant binding energy as a function of the energy from the Dirac point. To achieve the goal, we performed ARPES experiments with circularly polarized light and obtained dichroism data. We observe the OAM direction from the dichroic ARPES data and find the SAM structures. The results are consistent with the theoretical prediction, even in the high warping region.

$Bi_2Te_3$ single crystals were grown by a vacuum Bridgman technique.[15] ARPES measurement was performed at the beam line 7U of UVSOR-II and HiSOR BL9A. Samples were cleaved *in situ* at 15 K and the chamber pressure was better than $5 \times 10^{-11}$ Torr. Photoemission spectra were acquired with a hemispherical photoelectron analyzer (MBS A-1 at UVSOR-II and Scienta R4000 at HiSOR). We used left- and right-circularly polarized (LCP and RCP) 8-eV photons to avoid signal from the bulk states.[16] Circular polarization was achieved by using a $MgF_2$ $\lambda/4$ wave plate at UVSOR-II and by an elliptically polarized undulator at HiSOR. The energy resolution was about 12 meV and angular resolution was about 0.1°, which corresponds to a momentum resolution of 0.0016 Å$^{-1}$.

Figure 1(a) shows the experimental geometry and the surface Brillouin zone (BZ) of $Bi_2Te_3$. CPL comes in at 40° to the sample surface. Figures 1(b) and 1(c) are ARPES data acquired with LCP and RCP light, respectively. Dirac conelike dispersive bands of the TM states, especially the upper cone, are observed in the data. The Dirac point is located at ≈0.27 eV. Figure 1(d) plots the difference between LCP and RCP data. It shows a clear CD signal over a wide energy range. Surprisingly, the dichroism has very strong binding energy dependence. For example, in Fig. 1(d), the dichroism in the right upper Dirac bands switches from negative to positive as we move from the Dirac point toward the Fermi energy. This casts a contrast to the $Bi_2Se_3$ case where the dichroism has small energy dependence.[14]

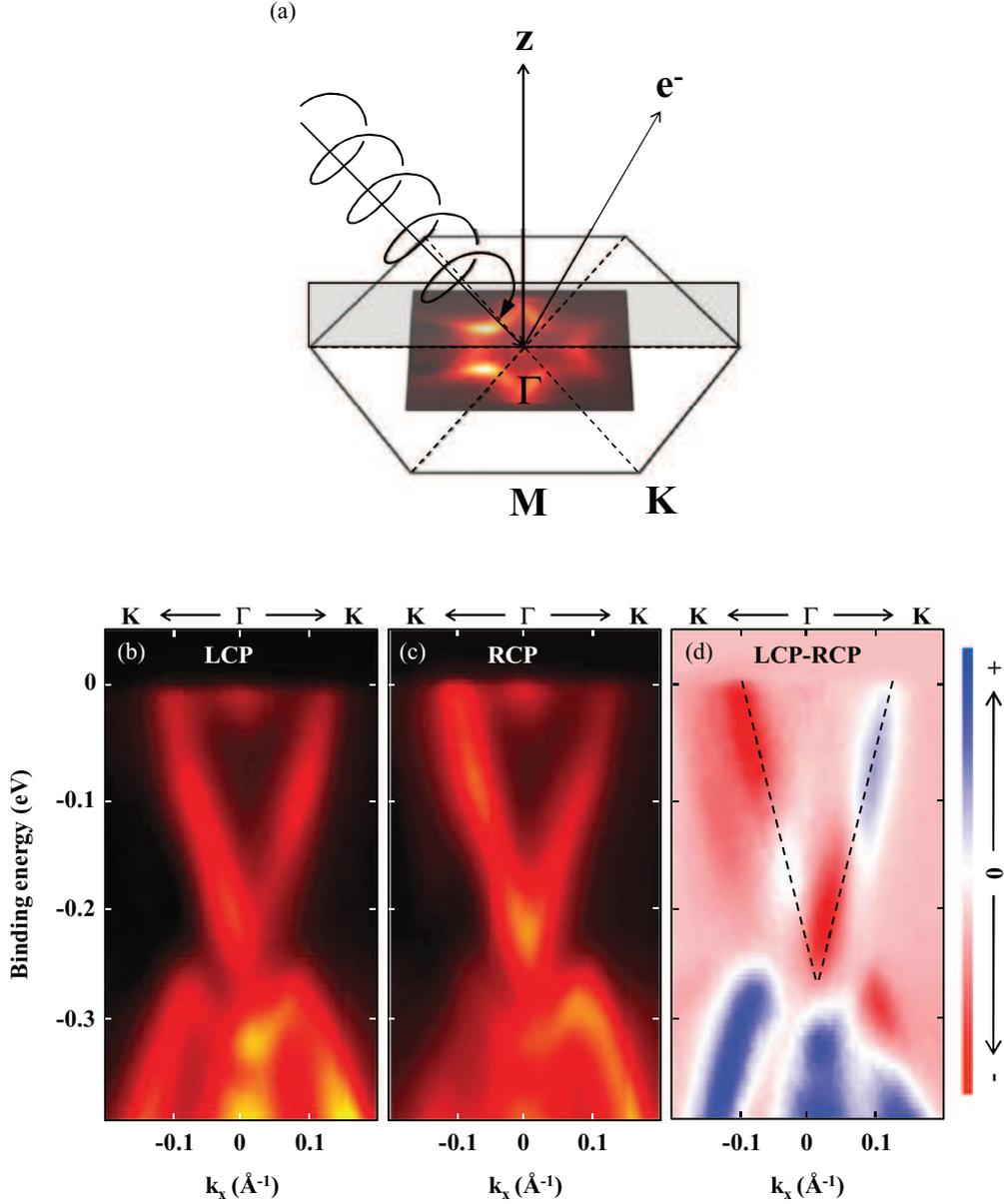

FIG. 1. (Color online) (a) Experimental geometry used in the study and Brillouin zone of $Bi_2Te_3$. Sample surface is parallel to the $xy$ plane. Circularly polarized photons (spiral arrow) come in in the $xz$ plane at an angle of 40° to the sample surface. (b) Γ-K cut taken with LCP. (c) Same cut taken with RCP. (d) Difference data taken with LCP and RCP.

One important difference between $Bi_2Se_3$ and $Bi_2Te_3$ is that the band warping effect is much more severe in the case of $Bi_2Te_3$. To see how such a warping effect plays a role in the dichroism, we need to inspect the CD in the momentum space at a constant binding energy. Figure 2(b) plots constant energy CD data taken with 50-meV steps between 0 and 250 meV as marked by the dashed lines in Fig. 2(a). The plots show that the sign of the CD on a constant energy contour has a binding energy dependence. Figure 2(c) shows CD data at 0-, 100-, and 250-meV binding energies. At 250 meV, which is close to the Dirac point, the shape of the constant energy contour is almost a circle. For 150-meV binding energy, the constant energy contour has a hexagonal shape. The Fermi contour at 0-meV binding energy has a strong warping effect, as manifested by its star shape. It is obvious that CD is also affected by the warping effect in $Bi_2Te_3$ and has a complex pattern.

One can consider various possibilities as the origin of the dichroism. First of all, it is known that CD in ARPES can result from the broken chirality in the experimental geometry. However, in such a case, the dichroism signal is expected to monotonically increase as the momentum moves away from the mirror plane (Γ-K line in this case) and cannot have the complex pattern shown in the leftmost panel in Fig. 2(c).[17] Therefore the most probable cause for the dichroism is the existence of OAM.[14,18] It was previously shown that CD is approximately proportional to the inner product of the light propagation and the OAM vectors.[14] For TIs, Berry's phase refers to the relative phases of the spin-up and spin-down components of the wave function. In such a case, it is interesting to note that CD is a direct measurement of the Berry's phase in the $k$ space, since the Berry's phase angle is precisely the angle of the spin measured from the $k_x$ axis of the Brillouin zone. A similar case has been reported for graphene.[19]

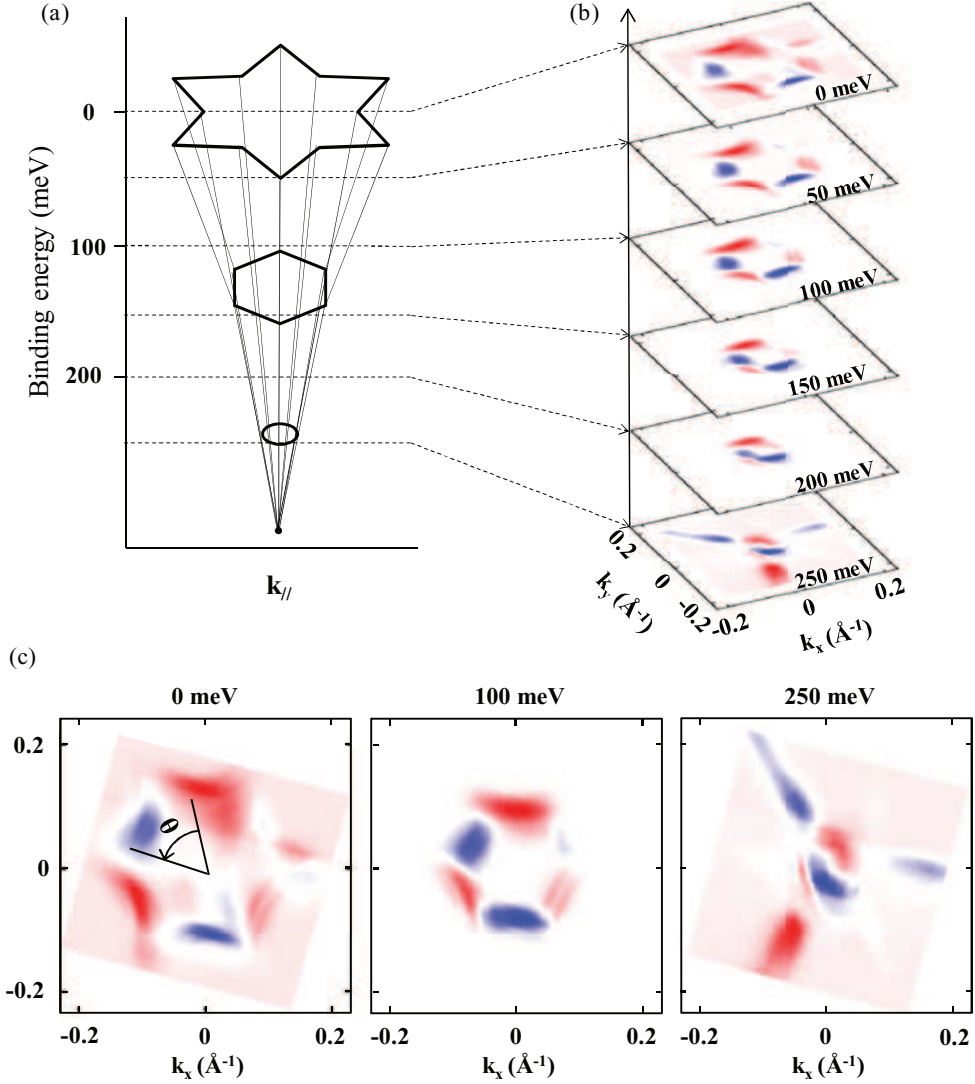

FIG. 2. (Color online) (a) Illustration of Dirac cone of $Bi_2Te_3$. (b) Dichroism in the momentum space at various binding energies. (c) Constant dichroism map at selected binding energies of 0, 100, and 250 meV.

CD along a constant energy contour is obtained and plotted in Fig. 3. It is plotted as a function of azimuthal angle $\theta$ as defined in Fig. 2(c). We define the CD intensity as the normalized difference $I_{CD} = (LCP - RCP)/(LCP + RCP)$. $I_{CD}$ becomes larger as the momentum goes away from the Dirac point and reaches the maximum value of about 40% at the Fermi energy. Such a large circular dichroism in ARPES is comparable to the previously reported value for $Bi_2Se_3$. Looking at the curves in Figs. 3(a)–3(f), it is evident that the behavior of $I_{CD}$ has very strong energy dependence. Near the Dirac point in Fig. 3(a), CD has an approximate form of $\sin\theta$. A strong coupling between OAM and SAM makes OAM to form chiral states as well. In an ideal case of chiral OAM state, the circular dichroism on a constant energy contour has a $\sin\theta$ form. Therefore our CD data shows that OAM has an almost ideal chiral structure near the Dirac point. Ideal chirality breaks down once the warping effect starts to come in. At 150-meV binding energy, which gives a hexagonal constant binding energy contour (weak warping effect region), $I_{CD}$ is close to a $\sin 3\theta$ function. By the time the energy reaches the Fermi energy, the warping effect is much stronger (strong warping effect region) and, as a result, the constant energy contour (Fermi surface in this case) is star-shaped. At this energy, $I_{CD}$ takes a complicated form and requires higher order sinusoidal functions.

As briefly mentioned above, such a complex CD pattern implies an equally complex OAM structure which originates from the warping effect. Then what is the OAM structure? We find that the CD pattern at 150 meV can be explained within the theoretical prediction.[12,13] That is, the warping effect is caused by higher order terms in the $kp$ perturbation when the contour size becomes larger (the band has a larger $k$ value). To describe the SAM structure, they discuss the in- and out-of-plane components separately. The in-plane component is chiral near the Dirac cone and turns into a wiggled chiral shape when the strength of the warping effect gets stronger. Meanwhile, the out-of-plane component is almost zero near the Dirac cone but becomes proportional to $\sin 3\theta$ as the strength of the warping effect increases.

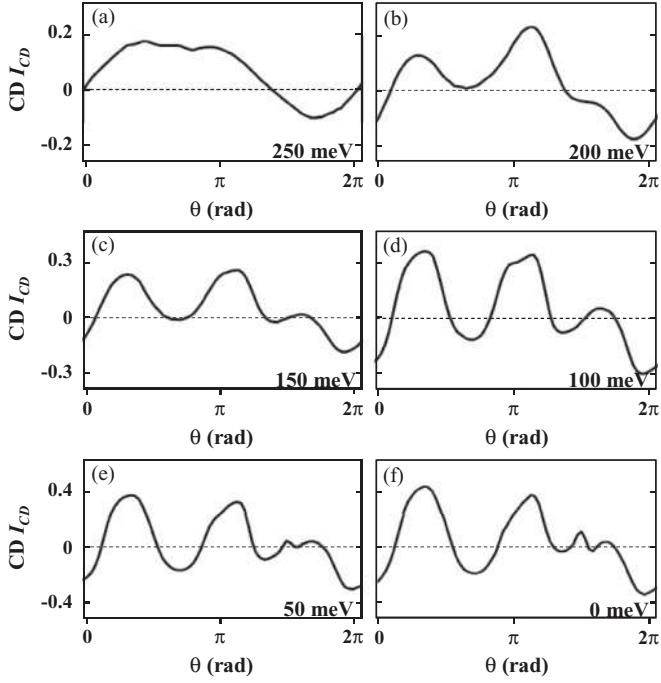

FIG. 3. Circular dichroism $I_{CD}$ as a function of $\theta$ at binding energies of (a) 250 meV, (b) 200 meV, (c) 150 meV, (d) 100 meV, (e) 50 meV, and (f) 0 meV. The definition for the azimuthal angle $\theta$ is shown in Fig. 2(c).

On the other hand, circular dichroism in ARPES comes from the OAM as discussed above, not from SAM. Therefore we need to incorporate this fact in the effective Hamiltonian to explain a complex CD pattern. We recently obtained an OAM-based effective Hamiltonian for Rashba-type band splitting.[20,21] The new Hamiltonian is given by

$$H_K = \alpha_K (\vec{E}_s \times \vec{k}) \cdot \vec{L}, \qquad (1)$$

where $\alpha_K$ is a constant that is proportional to the degree of asymmetric charge distribution for a given OAM, $\vec{E}_s$ the surface electric field, and $\vec{L}$ the OAM. The formalism for this new Hamiltonian simply replaces the SAM of the Rashba Hamiltonian by OAM, yet it can solve the energy scale problem in Rashba-type split bands.[22] Now we can write the $kp$ perturbation equation for the warped Dirac cone by using the OAM-based terms,

$$H(k) = E_0(k) + \upsilon_k(k_x L_y - k_y L_x) + \lambda_k/2(k_+^3 + k_-^3)L_z \\ + i\zeta(k_+^5 L_+ - k_-^5 L_-) + \cdots \qquad (2)$$

The Hamiltonian, except that the SAM is replaced by OAM, is the same as the previously reported one.[12,13] Then we can expect that the complex OAM pattern obtained from the CD can be explained in the same way.

To go beyond qualitative aspect of the CD to a quantitative analysis, we fit the CD data in Fig. 3 and plot the result in Fig. 4. Data near the Dirac point plotted in Fig. 4(a) can be fitted with a single $\sin\theta$ function, which indicates that the OAM has an almost ideal helicity near the Dirac point (250 meV), as schematically shown in the figure. Meanwhile, fitting the 100-meV data shown in Fig. 4(b) cannot be done with a single $\sin\theta$ but requires an additional term of $\sin 3\theta$, in the form of $A\sin\theta + B\sin 3\theta$. The magnitude of the $\sin\theta$ term is somewhat smaller than that of the near-Dirac-point data at 250 meV. Here, we note that the sign of the data changes five times (due to the $\sin 3\theta$ term), and the in-plane component alone cannot explain such a sign change because the chirality of the in-plane component cannot be reversed (that is, the in-plane component can change the sign change only once). Therefore, the $\sin 3\theta$ term comes from the out-of-plane component of the OAM. In fact, CPL comes in at an angle of 40° to the sample surface in our experimental geometry (Fig. 1) and thus the dichroism picks up the out-of-plane component. All things considered, our result is consistent with the prediction by Fu.[12] In addition, it is also compatible with the recently reported spin-resolved ARPES data.[23,24] Spin-resolved ARPES data show that the in-plane component has a chiral structure with an approximately constant magnitude, while the out-of-plane component changes in sign and gives the $\sin 3\theta$ term.

As noted above, CD at the Fermi surface has a more complicated shape, which implies that the OAM texture is also more complex. We first note that $I_{CD}$ at 0 meV has a double-peak structure near $\theta = 2\pi/3$ [arrow in Fig. 4(c)]. This feature has been observed in repeated measurements and thus is not from noise. It turns out that fitting such a structure requires an additional term of $\sin 6\theta$. This can be explained by the fact that higher order terms are needed in the $kp$ perturbation theory as the size of the momentum increases. In fact, the size of the constant energy contour of $Bi_2Te_3$ at the Fermi level is almost twice larger than that of $Bi_2Se_3$. The effect from higher order terms is discussed in a recent theoretical work by Basak et al.[13] on the spin texture of the strong warping case. They consider higher order terms to explain the nonorthogonality between the spin and the momentum. Nonorthogonality ($\delta$ in the report), expressed as a function of $\theta$, oscillates as $\sin 6\theta$. Therefore, the $\sin 6\theta$ modulation in our data may have come from the nonorthogonality of in-plane-direction OAM caused by the large Fermi surface. Besides, noting that the $\sin 6\theta$ modulation which is even function breaks the time-reversal symmetry, we expect that $\sin 6\theta$ modulation can be caused by extrinsic channels such as magnetic impurities or surface-bulk scattering.[16] It is interesting to note that bottom of the bulk conduction band exists in the strong warping effect region.

At this stage, it is also worth mentioning that sensitivity of CD to the in- and out-of-plane OAM can be quite different. As discussed above, the in-plane component of OAM at 100 meV has a chiral structure with an approximately constant magnitude. Constant magnitude of the in-plane component implies that the magnitude of the out-of-plane component is small (otherwise, the in-plane component would be more severely modulated). Yet, looking at the 100-meV data in Fig. 4(b), one finds that $\sin 3\theta$ (which comes from the out-of-plane component) is as large as the $\sin\theta$ term (in-plane component). The ratio between $\sin\theta$ and $\sin 3\theta$ obtained from the fitting is 1:1.2. Therefore, for the photon energy we used, CD seems to be more sensitive to the out-of-plane component than to the in-plane component. We expect that such a dichotomy in the sensitivity has photon energy dependence because the final state is modified as the photon energy changes.

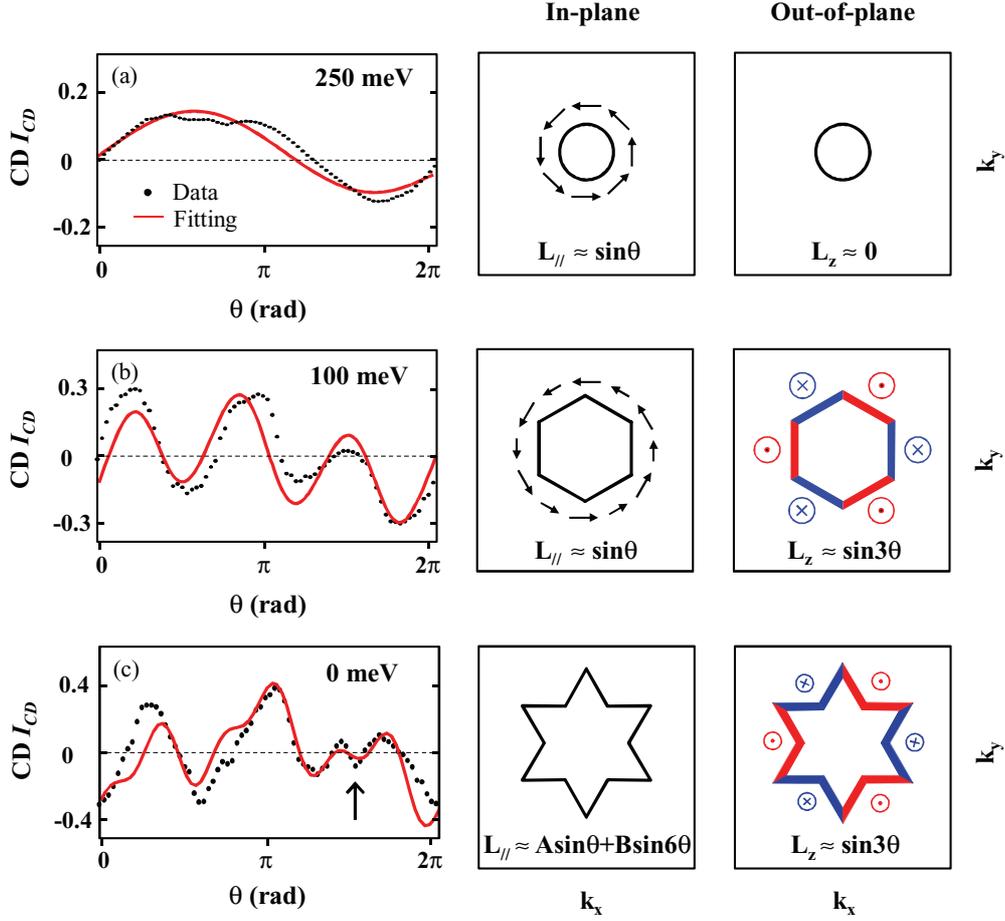

FIG. 4. (Color online) CD along the constant energy contour at (a) 0 meV, (b) 100 meV, and (c) 250 meV. Dots and lines represent the data and the fit, respectively. Also drawn are schematics of the in-plane (black arrow) and out-of-plane components of the OAM. For the out-of-plane components, the red segment of the contour has in-to-page OAM while blue does out-of-page, as shown by the symbols.

In conclusion, we investigated the OAM of $Bi_2Te_3$ by using ARPES with circularly polarized light. Similar to CD results from $Bi_2Se_3$, our data show very strong CD. We find that CD due to OAM is severely affected by the warping effect in the band structure. The OAM close to the Dirac point has an ideal chiral structure ($\sin\theta$) without the out-of-plane component. The CD shows a $\sin 3\theta$ effect in the weak warping effect region around 150-meV binding energy. Such $\theta$ dependence is compatible with the theoretically predicted spin structure for the case when the out-of-plane OAM component has a strong contribution to the $\theta$ dependence. In addition, CD gets an extra $\sin 6\theta$ term from modulation in the in-plane OAM component when the warping effect becomes very strong near the Fermi energy. This result is consistent with theoretical prediction, which considers higher order terms in our modified $kp$ perturbation theory that incorporates OAM.

The authors acknowledge helpful discussions with S. R. Park. This work was supported by the NRF (Contract No. 20090080739) and the KICOS under Grant No. K20602000008, performed as a part of a joint studies program of the Institute for Molecular Science in 2010 and with the approval of the Proposal Assessing Committee of HSRC (Proposal No. 10-A-57). This work was performed by the use of the UVSOR Facility Program (BL7U, 2011) of the Institute for Molecular Science.

*changyoung@yonsei.ac.kr
[1] L. Fu, C. L. Kane, and E. J. Mele, Phys. Rev. Lett. **98**, 106803 (2007).
[2] L. Fu and C. L. Kane, Phys. Rev. B **76**, 045302 (2007).
[3] J. E. Moore and L. Balents, Phys. Rev. B **75**, 121306 (2007).
[4] X. L. Qi, T. L. Hughes, and S. C. Zhang, Phys. Rev. B **78**, 195424 (2008).

[5] J. C. Y. Teo, L. Fu, and C. L. Kane, Phys. Rev. B **78**, 045426 (2008).
[6] H. J. Zhang, C. X. Liu, X. L. Qi, X. Dai, Z. Fang, and S. C. Zhang, Nat. Phys. **5**, 438 (2009).
[7] D. Hsieh, Y. Xia, L. Wray, D. Qian, A. Pal, J. H. Dil, J. Osterwalder, F. Meier, G. Bihlmayer, C. L. Kane, Y. S. Hor, R. J. Cava, and M. Z. Hasan, Science **323**, 919 (2009).
[8] T. Zhang, P. Cheng, X. Chen, J. F. Jia, X. Ma, K. He, L. Wang, H. Zhang, X. Dai, Z. Fang, X. Xie, and Q. K. Xue, Phys. Rev. Lett. **103**, 266803 (2009).
[9] Z. Alpichshev, J. G. Analytis, J. H. Chu, I. R. Fisher, Y. L. Chen, Z. X. Shen, A. Fang, and A. Kapitulnik, Phys. Rev. Lett. **104**, 016401 (2010).
[10] Y. L. Chen, J. G. Analytis, J.-H. Chu, Z. K. Liu, S.-K. Mo, X. L. Qi, H. J. Zhang, D. H. Lu, X. Dai, Z. Fang, S. C. Zhang, I. R. Fisher, Z. Hussain, and Z.-X. Shen, Science **325**, 178 (2009).
[11] K. Kuroda, M. Arita, K. Miyamoto, M. Ye, J. Jiang, A. Kimura, E. E. Krasovskii, E. V. Chulkov, H. Iwasawa, T. Okuda, K. Shimada, Y. Ueda, H. Namatame, and M. Taniguchi, Phys. Rev. Lett. **105**, 076802 (2010).
[12] L. Fu, Phys. Rev. Lett. **103**, 266801 (2009).
[13] S. Basak, H. Lin, L. A. Wray, S.-Y. Xu, L. Fu, M. Z. Hasan, and A. Bansil, Phys. Rev. B **84**, 121401(R) (2011).
[14] S. R. Park, J. Han, C. Kim, Y. Y. Koh, C. Kim, H. Lee, H. J. Choi, J. H. Han, K. D. Lee, N. J. Hur, M. Arita, K. Shimada, H. Namatame, and M. Taniguchi, e-print arXiv:1103.0805.
[15] Y. S. Hor, A. Richardella, P. Roushan, Y. Xia, J. G. Checkelsky, A. Yazdani, M. Z. Hasan, N. P. Ong, and R. J. Cava, Phys. Rev. B **79**, 195208 (2009).
[16] S. R. Park, W. S. Jung, C. Kim, D. J. Song, C. Kim, S. Kimura, K. D. Lee, and N. Hur, Phys. Rev. B **81**, 041405(R) (2010).
[17] R. L. Dubs, S. N. Dixit, and V. McKoy, Phys. Rev. Lett. **54**, 1249 (1985).
[18] C. H. Kim, J.-H. Park, J. W. Rhim, B. Y. Kim, J. Yu, M. Arita, K. Shimada, H. Namatame, M. Taniguchi, C. Kim, and J. H. Han, e-print arXiv:1107.3285.
[19] Y. Liu, G. Bian, T. Miller, and T.-C. Chiang, Phys. Rev. Lett. **107**, 166803 (2011).
[20] B. Y. Kim *et al.*, unpublished.
[21] S. R. Park, C. H. Kim, J. Yu, J. H. Han, and C. Kim, Phys. Rev. Lett. **107**, 156803 (2011).
[22] P. D. C. King, R. C. Hatch, M. Bianchi, R. Ovsyannikov, C. Lupulescu, G. Landolt, B. Slomski, J. H. Dil, D. Guan, J. L. Mi, E. D. L. Rienks, J. Fink, A. Lindblad, S. Svensson, S. Bao, G. Balakrishnan, B. B. Iversen, J. Osterwalder, W. Eberhardt, F. Baumberger, and Ph. Hofmann, Phys. Rev. Lett. **107**, 096802 (2011).
[23] S. Souma, K. Kosaka, T. Sato, M. Komatsu, A. Takayama, T. Takahashi, M. Kriener, K. Segawa, and Y. Ando, Phys. Rev. Lett. 106, 216803 (2011).
[24] Z. H. Pan, E. Vescovo, A. V. Fedorov, D. Gardner, Y. S. Lee, S. Chu, G. D. Gu, and T. Valla, Phys. Rev. Lett. 106, 257004 (2011).